\input epsf
\documentstyle[12pt]{article}
\begin{document}
\begin{titlepage}
\setcounter{page}{0}
\rightline{YERPHI-1449(19)-95}
\rightline{TECHNION-PH-96-1}
\rightline{hep-ph/9603318}
\vspace{2cm}
\begin{center}
{\Large  CP-Violation in the Decay $b \rightarrow s\gamma$
   in the Left-Right Symmetric Model}
\vspace{1cm}

{\large H. M. Asatrian$^{a}$, A. N. Ioannissian$^{a,b}$} \\

\vspace{1cm}

{\em $a$ ~~Yerevan Physics Institute, Alikhanyan Br. 2, Yerevan, Armenia}\\
{\em $b$ ~~Dept. of Physics, Technion - Israel Inst. of Tech., Haifa, Israel}\\
{\em e-mail: asatryan@vx1.yerphi.am, ioannissyan@vxc.yerphi.am}\\
\end{center}
\vspace{5mm}

\centerline{{\bf{Abstract}}}
The direct CP-violation in the left-right symmetric
$SU(2) \times SU(2) \times U(1)$ model is investigated for
the decay $b \rightarrow s\gamma$. The calculated CP-asymmetry
for the wide range of parameters can be larger than in standard model
and can have an opposite sign.

\vfill
\centerline{\large Yerevan Physics Institute}
\centerline{\large Yerevan 1995}

\end{titlepage}

\newpage

The experimental and theoretical investigation of the decay
$b \rightarrow s \gamma$ can give a sign for a new physics in the
TeV region \cite{1}. This decay has been extensively studied  during the
last years. The first experimental evidence was obtained at  CLEO for the
exclusive decay $\bar{B} \rightarrow K^{*} \gamma$ \cite{2}. The decay
$b \rightarrow s \gamma$ has been investigated theoretically for the
standard model and its  extensions \cite{3}-\cite{11}.
CP-violation in $B-\bar{B}$ system was considered
in \cite{12}.

In this paper we  consider the CP asymmetry in the decay
$b \rightarrow s \gamma$ for
the left-right symmetric $SU(2) \times SU(2) \times U(1)$ model,
which is one of the simplest extensions of the standard model.
Calculated value of CP-asymmetry
for some range of parameters of the model (the mass of the right W-boson,
the ratio of two Higgs doublet vacuum expectation values, phases and
mixing angles) is almost 2 times larger than in standard
model and can have an opposite sign,
while the decay rate is almost the same.

The Lagrangian of interaction of quarks with scalar and
$SU(2) \times SU(2) \times U(1)$ gauge fields has the following form:
\begin{eqnarray}
\nonumber
L=(A_{ik} \bar{\Psi}_{Li}\Phi\Psi_{Ri}+
 B_{ik} \bar{\Psi}_{Li} \tilde{\Phi}\Psi_{Ri} +c.c)+ \\
ig_{R}\bar{\Psi}_{Ri}\hat{W}_{R}^{a}\sigma_{a}\Psi_{Ri}  %(1)
+ ig_{L}\bar{\Psi}_{Li}\hat{W}_{L}^{a}\sigma_{a}\Psi_{Li}
\end{eqnarray}
where i,k=1,2,3 and
\begin{displaymath}
\Phi=\left(\begin{array}{cc}
\eta^e &\xi^+\\
\eta^- & -\xi^{o*}
\end{array}\right) \hspace{0.3cm}
\tilde{\Phi}= \left(\begin{array}{cc}
\xi^o &\eta^+\\
\xi^- & -\eta^{o*}
\end{array}\right)\hspace{0.3cm}
\Psi_{i,L,R}=\left(\begin{array}{c}
U\\D \end{array}\right)_{i,L,R}
\end{displaymath}
The symmetry $SU(2) \times SU(2) \times U(1)$ can be broken
to $SU(2)_{L} \times U(1)$ by means of vacuum expectation
values (vev) of doublet or triplet fields \cite{9,10,11}. As for
$SU(2) \times U(1)$ symmetry breaking we assume that it takes place
when the scalar field $\Phi$ acquires the vev:
\begin{equation}
\Phi=
\left( \begin{array}{cc}
k & 0\\                                   %(2)
0 & -e^{i\delta} k'\\
\end{array} \right),
\end{equation}
The interaction of quark charged current with W gauge
boson and charged Higgs fields has the form:
\begin{eqnarray}
L^{ch}&=&\frac{1}{\sqrt{2}}\left(\bar{u},\bar{c},\bar{t}\right)\left[
\hat{W_1}^+\left[-g_L\cos\beta K_L P_--g_R\sin\beta  e^{i\delta }K_RP_+
\right]\right.+\nonumber \\
&+&\varphi ^+\frac{g_L }{\sqrt{2}M_{W_L}}\left[ \left(-\tan2\theta K_LM_d+
e^{i\delta }\frac{1}{\cos2\theta}M_uK_R\right)P_++\right.\\
&+&\left.\left.\left( \tan2\theta                                      %(3)
 M_uK_L-e^{i\delta }\frac{1}{\cos2\theta}K_RM_d\right)
P_-\right]\right]\left(\begin{array}{l}
d\\s\\b \end{array}\right)\nonumber
\end{eqnarray}
where $W_1$ is the "light" (predominantly 
left-handed) charge gauge boson
and $\beta$ is the mixing angle between left and right W - bosons,
\begin{displaymath}
\tan2\beta  =2\sin2\theta
\frac{g_R}{g_L}
\frac{M^2_{W_L}}{M^2_{W_R}}
\hspace{0.4cm} \tan\theta=-\frac{k'}{k}
\end{displaymath}
$K_{L}$ and $K_{R}$ are Kobayashi-Maskawa mixing matrices
for left and right charged currents respectively,
$P{\pm}=(1\pm \gamma_{5})/2$,
\begin{displaymath}
M_u=\left(
\begin{array}{lll}
m_u & 0&0\\
0  & m_c&0\\
0&0&m_t
\end{array}\right),
\quad \quad
M_d=\left(
\begin{array}{ccc}
m_d & 0&0\\
0  & m_s&0\\
0&0&m_b
\end{array}\right)
\end{displaymath}
The matrices $K_{L}$ and $K_{R}$ can be expressed in such
a form where $K_{L}$
has only one complex phase and $K_{R}$ has five complex phases
\cite{13}. In (3)
we omit the term, connected with the interaction with heavy (predominantly right)
W- boson, which is not relevant for $b \rightarrow s \gamma$ decay.

The effective lagrangian for  $b \rightarrow s \gamma$ decay
has the following form \cite{9,10,11}:
\begin{eqnarray}
\nonumber
H_{b\rightarrow s\gamma} & = & -\frac{e}{16\pi^2}
\frac{2G_F}{\sqrt{2}}m_b\left(K_{ts}^{L*}K_{tb}^L A_{s\gamma }^{W_L}O_7^L
+ e^{i\delta}K_{ts}^{L*}K_{tb}^R \beta\frac{m_t}{m_b^*}
A_{s\gamma}^{W_{LR}}O_7^L \right.+ \\                                    %(4)
& + & e^{i\delta}K_{ts}^{L*}K_{tb}^R \frac{\sin 2\theta }{\cos^22\theta}
\frac{m_t}{m_b^*}A_{s\gamma }^{\varphi ^+}O_7^L+
\frac{m_s}{m_b} K_{ts}^{L*}K_{tb}^L A_{s\gamma}^{W_{L}}O_7^R \\
\nonumber
&+&\left. e^{-i\delta}K_{ts}^{R*}K_{tb}^L \beta\frac{m_t}{m_b^*}
A_{s\gamma }^{W_{RL}}O_7^R+
e^{-i\delta}K_{ts}^{R*}K_{tb}^L \frac{\sin 2\theta }{\cos^22\theta}
\frac{m_t}{m_b^*}A_{s\gamma }^{\varphi ^+}O_7^R\right)
\end {eqnarray}
where
\begin{eqnarray}
\nonumber
&&O_7^{L,R}=\bar{u}_s \sigma^{\mu\delta}(1\pm \gamma
_5)u_b F_{\mu \nu }, \hspace{0.4cm}
O_8^{L,R}=\bar{u}_s\sigma^{\mu \delta  }(1\pm \gamma         %(5)
_5)u_b G_{\mu \nu }\\
&&A_{s\gamma }^{W_L}=\eta
^{\frac{16}{23}}\left[{\cal A}^{W_L}_{s\gamma}+\frac{8}{3}
{\cal A}^{W_L}_{sg} \left( \eta^{-\frac{2}{23}}-1
\right)-\frac{232}{513}\left( \eta^{-\frac{19}{23}}-1
\right)\right]\nonumber \\
&&A_{s\gamma }^{W_{LR}}=\eta
^{\frac{16}{23}}\left[{\cal A}^{W_{LR}}_{s\gamma}+\frac{8}{3}
{\cal A}^{W_{LR}}_{sg} \left( \eta^{-\frac{2}{23}}-1
\right)\right] \\
\nonumber
&&A_{s\gamma }^{\varphi^ +}=\eta^{\frac{16}{23}}
\left[{\cal A}^{\varphi^ +}_{s\gamma}+\frac{8}{3}{\cal A}^{\varphi^{+}}_{sg}
\left( \eta^{-\frac{2}{23}}-1
\right)\right]
\end{eqnarray}
where $\eta =\frac{\alpha_{s}(m_{W})}{\alpha_{s}(m_{b})}$
and the functions
${\cal A}_{s\gamma}^{W_{L}}$,
${\cal A}_{s\gamma}^{W_{L,R}}$,
${\cal A}_{s\gamma}^{\phi^{+}}$
${\cal A}_{sg}^{W_{L}}$,
${\cal A}_{sg}^{W_{L,R}}$,
${\cal A}_{sg}^{\phi^{+}}$
was presented in  \cite{9,10,11,14,15,16}:
\begin{eqnarray}
\nonumber
&&{\cal A}_{s\gamma}^{W_{L}}=Q_tF_1(x)+G_1(x), \hspace{0.5cm}
{\cal A}_{sg}^{W_{L}}=F_1(x) \\
&&{\cal A}_{s\gamma}^{W_{L,R}}=Q_tF_2(x)+G_2(x), \hspace{0.2cm} %(6)
{\cal A}_{sg}^{W_{L,R}}=F_2(x) \\
\nonumber
&&{\cal A}_{s\gamma}^{\phi^{+}}=Q_tF_3(y)+G_3(y), \hspace{0.6cm}
{\cal A}_{sg}^{\phi^{+}}=F_3(y)
\end{eqnarray}
and
\begin{eqnarray}
\nonumber
&&F_1(x)=-\frac{1}{(1-x)^4}\left [\frac{x^4}{4}-\frac{3}{2}x^3+\frac{3}{4}
x^2+\frac{x}{2}+\frac{3}{2}x^2\log(x)\right ] \\
\nonumber
&&G_1(x)=-\frac{1}{(1-x)^4}\left [\frac{x^4}{2}+\frac{3}{4}x^3-\frac{3}{2}  %(7)
x^2+\frac{x}{4}-\frac{3}{2}x^3\log(x)\right ] \\
&&F_2(x)=-\frac{1}{(1-x)^3}\left [-\frac{x^3}{2}-\frac{3}{2}
x+2+3x\log(x)\right ] \\
\nonumber
&&G_2(x)=-\frac{1}{(1-x)^3}\left [-\frac{x^3}{2}+6x^2-\frac{15}{2}
x+2-3x^2\log(x)\right ] \\
\nonumber
&&F_3(y)=-\frac{1}{(1-y)^3}\left [-\frac{y^3}{2}+2y^2-\frac{3}{2}
y-y\log(y)\right ] \\
\nonumber
&&G_3(y)=-\frac{1}{(1-y)^3}\left [-\frac{y^3}{2}+\frac{1}{2}
y+y^2\log(y)\right ]
\end{eqnarray}
where $Q_t=2/3$ is the electric charge of the top quark, $x=m_t^2/m_W^2$,
$y=m_t^2/m_{\varphi^+}^2$.
The direct CP- asymmetry     for  $b \rightarrow s \gamma$ decay arises only
when one take into account the final state interaction effects, when the
absorptive parts arise.

Absorptive  parts of the decay amplitude arise from rescattering
$b\rightarrow su\bar{u} \rightarrow s\gamma$,
$b\rightarrow sc\bar{c} \rightarrow s\gamma$,
$b\rightarrow sg \rightarrow s\gamma$:
\begin{eqnarray}
\nonumber &&H_{b\rightarrow s\gamma}^{absorbt}=i\frac{e}{16\pi^2}
\frac{2G_F}{\sqrt{2}}m_b
\left\{ K_{ts}^{L*}K_{tb}^L
A_{sg}^{W_L}O_8^L t_{sg\rightarrow s\gamma }
+\sum_{q=u,c}K_{qs}^{L*}K_{qb}^L A_{sq\bar{q}}^{W_L}
t_{sg\rightarrow s\gamma} +\right.\\
&&+ e^{i\delta}K_{ts}^{L*}K_{tb}^R \beta\frac{m_t}{m_b^*}
A_{sg}^{W_{RL}}O_8^L t_{sg\rightarrow s\gamma }
+ e^{i\delta}K_{ts}^{L*}K_{tb}^R \frac{\sin 2\theta }{\cos^22\theta }  %(8)
\frac{m_t}{m_b^*}
A_{sg }^{\varphi^+}O_8^L t_{sg\rightarrow s\gamma }+\\                %
\nonumber &&+e^{-i\delta}K_{ts}^{R*}K_{tb}^L \beta\frac{m_t}{m_b^*}
A_{sg }^{W_{RL}}O_8^R t_{sg\rightarrow s\gamma } +
e^{-i\delta}\left.K_{ts}^{R*}K_{tb}^L \frac{\sin 2\theta }{\cos^22\theta }
\frac{m_t}{m_b^*}
A_{sg }^{\varphi^+}O_8^R t_{sg\rightarrow s\gamma
}\right\}
\end{eqnarray}
where
\begin{equation}
A_{sg}^{\varphi^ +}=  \eta                                        %(9)
^{\frac{14}{23}}{\cal A}^{\varphi^ +}_{sg}, \hspace{0.2cm}
A_{sg}^{W_L}=\eta^{\frac{14}{23}}\left[{\cal A}^{W_L}_{sg}
-0.1687 \right], \hspace{0.2cm}
A_{sg}^{W_{LR}}=\eta^{\frac{14}{23}}{\cal A}_{sg}^{W_{LR}}
\end{equation}
In the standard model only for the rescattering
$b\rightarrow su\bar{u} \rightarrow s\gamma$ and
$b\rightarrow sc\bar{c} \rightarrow s\gamma$ one obtains
nonegligible contribution \cite{14}. For the two Higgs doublet extension
of the standard model or left-right symmetric model the
rescattering $b\rightarrow sg \rightarrow s\gamma$ also must
be taken into consideration. Taking into account the standard
model result \cite{14} we obtain the following result for the absorptive
part of the decay
$b \rightarrow s \gamma$ in left-right symmetric model:
\begin{eqnarray}
\nonumber &&H_{b\rightarrow s\gamma}^{absorbt}\simeq
-i\frac{e}{16\pi^2}\frac{2G_F\alpha_s}{\sqrt{2}} m_{b}
\{O_7^L(\frac{2}{9}(K^{L_*}_{ts}K^L_{tb} A^{W_L}_{sg}+
e^{i\delta}K^{L^*}_{ts}K^L_{tb}
\frac{K^R_{tb}}{K^L_{tb}}A^R_{sg})+\\
&&+\frac{1}{4}(K^{L^*}_{us}K_{ub}^L                   %(10)
+0.12K^{L^*}_{cs}K^L_{cb})c_1)+
O_7^R\frac{2}{9}e^{-i\delta}K^{L^*}_{ts}K^L_{tb}
\frac{K^R_{ts}}{K^L_{ts}}
A^R_{sg}\}
\end{eqnarray}
where
\begin{eqnarray}
A^{R}_{sg} \equiv \beta\frac{m_{t}}{m_b^*}A_{sg}^{W_{LR}}+   % (11)
\frac{m_{t}}{m_b^*}\frac{\sin2\theta}{\cos^{2}2\theta}A^{\varphi^{+}}_{sg}
\end{eqnarray}
We note that our result is different from the one obtained in  \cite{15}:
the contribution of the rescattering
$b\rightarrow sg \rightarrow s\gamma$
differs from that in \cite{15} by the factor 2/9.
The CP-asymmetry for the decay $b \rightarrow s\gamma$
and $\bar{b} \rightarrow \bar{s}\gamma$ is defined
as:
\begin{equation}
a_{cp}=\frac{\Gamma(\bar{b} \rightarrow \bar{s}\gamma)-
\Gamma(b \rightarrow s\gamma)}
{\Gamma(\bar{b} \rightarrow \bar{s}\gamma)+
\Gamma(b \rightarrow s\gamma)}                                 %(12)
\end{equation}
The resulting CP-asymmetry is equal to:
\begin{eqnarray}
\nonumber
&&a_{cp}=\frac{2\alpha
_s}{(|C_7^L|^2+|C_7^R|^2) v_t^*v_t}
\left\{({\rm Im} v^*_t v_u+
0.12{\rm Im}v^*_tv_c)\times \right.\\                                 %(13)
&&\times (A^{W_{L}}_{s\gamma }+
H\cos \alpha  A^R_{s\gamma })\frac{c_1}{4}-({\rm Re} v^*_t v_u+
0.12{\rm Re} v^*_t v_c)\times \\
\nonumber
&&\left.\times A^R_{s\gamma }H \sin \alpha
\frac{c_1}{4}+
\frac{2}{9}H \sin \alpha v^*_t v_t(A_{s\gamma}^{W_{L}}A^R_{sg}-
A^R_{s\gamma }A_{sg}^{W_{L}})\right\}
\end{eqnarray}
where
\begin{eqnarray}
\nonumber
&&He^{i\alpha  }\equiv e^{i\delta }\frac{K_{Rtb}}{K_{Ltb}},
\hspace{0.2cm}A^{R}_{s \gamma} \equiv \beta\frac{m_{t}}
{m_b^*}A_{s\gamma}^{W_{LR}}+
\frac{m_{t}}{m_b^*}\frac{\sin2\theta}{\cos^{2}2\theta}
A^{\varphi^{+}}_{s\gamma}\\
&&v_t\equiv K_{Lts}^*K_{Ltb},\quad \quad v_c\equiv
K_{Lcs}^*K_{Lcb},\quad \quad v_u\equiv K_{Lus}^*K_{Lub}\\
\nonumber
&&C_7^L= A_{s\gamma }^{W_L}+\left(e^{i\delta}\frac{K^R_{tb}}{K^L_{tb}}
\right)
A^R_{s\gamma },\quad C_7^R= e^{-i\delta
}\frac{K^{R^*}_{ts}}{K^{L^*}_{ts}} A^R_{s\gamma }                    %(14)
\end{eqnarray}
In the numerical results we take $\alpha_s=0.24$, $c_{1}\simeq 1.1$,
$m_t=175$GeV,
$m_b=4.5GeV$, $m_b^*\equiv m_b(m_t)=3GeV$ \cite{10}.
The CP-asymmetry  $a_{cp}$ depends on the parameters of
Kobayashi-Maskawa mixing matrix in Wolfensteins parametrization:
$\lambda$=0.221, $\eta$, $\rho$ \cite{14,17,18,19}, and also
it depends on the parameters
of left-right symmetric  model: $\alpha$, $\tan2\theta$, $M_{W_{R}}$,
$M_{\varphi^{+}}$, H. We assume that $|H|=|K^R_{tb}/K^L_{tb}|\simeq 1$.
For the fixed masses we vary the remaining parameters and obtain
the allowed region
of $a_{cp}$ values. We take into account that in the left-right symmetric
model the decay rate cannot differ sufficiently from those obtained in
the standard  model. The point is that the standard model predictions
for decay rate are reasonable agreement with experiment.
We present in Fig 1,2,3  the minimal and maximal values of the asymmetry for
the various $M_{W_{R}}$, $M_{\varphi^{+}}$  and $0<\tan2\theta<3.5$
(the maximal and minimal values of $a_{CP}$ are symmetric under changing
of sign of $\tan2\theta$).
It is easy to see that the $a_{CP}$ can be about 2 times larger than  the
maximal value of asymmetry in the standard model. For
the masses of right sector $\geq$5TeV the asymmetry is rather sensitive on
changes of the Higgs boson mass than the right W- boson mass.
In contrary to the standard model, where the asymmetry have a negative
sign, in our case the sign of the asymmetry can also be positive.
The results for $a_{CP}$ in Fig 1,2,3  are obtained under the
assumption that the decay rate in left-right symmetric model can differ from
the standard model predictions no more than
$\Delta=10\%$. Our calculations show that the results for maximal and
minimal values of $a_{CP}$ practically does not change when we
vary $\Delta$ from 10\% to 50\%.  This fact
is illustrated in Fig 4, where we compare the results for $\Delta$=10\%
and $\Delta$=50\% for masses $M_{W_{R}}=M_{\varphi^{+}}$=10TeV.

In table we present the $a_{CP}$ minimal and maximal values in left-right
symmetric model for some values of the parameters
$M_{W_{R}}$, $M_{\varphi^{+}}$,  $\tan2\theta$ and for
$\rho$, $\eta$ "best fit" \cite{19}: $(\rho,\eta)=
(-0.05,0.37)$. For $\eta=0.37$ the standard model prediction for
$a_{CP}$ is -0.64\%. For the same $\eta$ the value of $a_{CP}$
in left-right symmetric model can be almost 2.5 times larger.

In conclusion, we have calculated the CP asymmetry in the decay
$b\rightarrow s \gamma$ for the left-right symmetric model
$SU(2) \times SU(2) \times U(1)$.
We have shown that the CP-asymmetry
for the reasonable range of parameters can be
larger than in standard model and can have an opposite sign.

The research described in this publication was made possible in part
by Grant N MVU000 from the International Science Foundation. A. I. has been
supported by Lady Davis Trustship.
\newpage
\begin{center}
TABLE   \\
\vspace{1cm}
\begin{tabular}{|c|c|c|c|}  \hline
  & $\tan2\theta$=1 & $\tan2\theta$=2 & $\tan2\theta$=3  \\ \hline
$M_{W_{R}}$=1.5TeV, $M_{\varphi^{+}}$=10TeV &$(-1.51 \div 0.33)\%$ &
$(-0.76 \div  0.16)\%$ & -  \\ \hline
$M_{W_{R}}$=10TeV, $M_{\varphi^{+}}$=1.5TeV & - &
- & -  \\ \hline
$M_{W_{R}}$=10TeV, $M_{\varphi^{+}}$=10TeV &$(-0.90 \div -0.37)\%$ &
$(-1.27 \div 0.25)\%$ &$(-0.39 \div 0.41)\%$  \\ \hline
$M_{W_{R}}$=20TeV, $M_{\varphi^{+}}$=20TeV &$(-0.73 \div-0.55)\%$ &
$(-0.93 \div -0.35)\%$ &$(-1.21 \div -0.06)\%$  \\ \hline
$M_{W_{R}}$=50TeV, $M_{\varphi^{+}}$=10TeV &$(-0.93 \div-0.35)\%$ &
$(-1.31 \div 0.28)\%$ &$(-0.48 \div 0.44)\%$  \\ \hline
$M_{W_{R}}$=10TeV, $M_{\varphi^{+}}$=50TeV &$(-0.65 \div-0.63)\%$ &
$(-0.67 \div-0.61)\%$ &$(-0.74 \div -0.54)\%$  \\ \hline
$M_{W_{R}}$=50TeV, $M_{\varphi^{+}}$=50TeV &$(-0.66 \div-0.62)\%$ &
$(-0.70 \div-0.58)\%$ &$(-0.77 \div -0.51)\%$  \\ \hline
\end{tabular} \\
\vspace{1cm}
The minimal and maximal values of $a_{CP}$ for $\rho$=-0.05, $\eta$=0.37
and some values of $M_{W_{R}}$,
$M_{\varphi^{+}}$ and $\tan2\theta$. \\
\end{center}

\newpage
\vspace{2cm}
\begin{center}
{\large Figure Captions}\\
\end{center}

Fig1. The maximal and minimal values of $a_{CP}$ in \% for $M_{W_{R}}$=5TeV,
$M_{\varphi^{+}}$=5TeV (curves 1 and 2);
for $M_{W_{R}}$=10TeV,
$M_{\varphi^{+}}$=10TeV (curves 3 and 4);
for $M_{W_{R}}$=20TeV,
$M_{\varphi^{+}}$=20TeV (curves 5 and 6);
for $M_{W_{R}}$=50TeV,
$M_{\varphi^{+}}$=50TeV (curves 7 and 8).\\

Fig2. The maximal and minimal values of $a_{CP}$ in \% for $M_{W_{R}}$=1.5TeV and
$M_{\varphi^{+}}$=10TeV (curves 1 and 2);
$M_{W_{R}}$=5TeV and
$M_{\varphi^{+}}$=10TeV (curves 3 and 4);
$M_{W_{R}}$=10TeV and
$M_{\varphi^{+}}$=10TeV (curves 5 and 6);
$M_{W_{R}}$=20TeV and
$M_{\varphi^{+}}$=10TeV (curves 7 and 8).\\

Fig3. The maximal and minimal values of $a_{CP}$ in \% for $M_{W_{R}}$=10TeV and
$M_{\varphi^{+}}$=1.5TeV (curves 1 and 2);
$M_{W_{R}}$=10TeV and
$M_{\varphi^{+}}$=5TeV (curves 3 and 4);
$M_{W_{R}}$=10TeV and
$M_{\varphi^{+}}$=10TeV (curves 5 and 6);
$M_{W_{R}}$=10TeV and
$M_{\varphi^{+}}$=20TeV (curves 7 and 8).\\

Fig4. The maximal and minimal values of $a_{CP}$ in \% for $M_{W_{R}}$=10TeV,
$M_{\varphi^{+}}$=10TeV and allowed difference from standard model
10\% (curves 1 and 2);$M_{W_{R}}$=10TeV,
$M_{\varphi^{+}}$=10TeV and allowed difference from standard model
50\% (curves 3 and 4).

\newpage
\begin{figure}[htb]
\epsfxsize=15cm
\epsfysize=10cm
\mbox{\hskip 0cm}\epsfbox{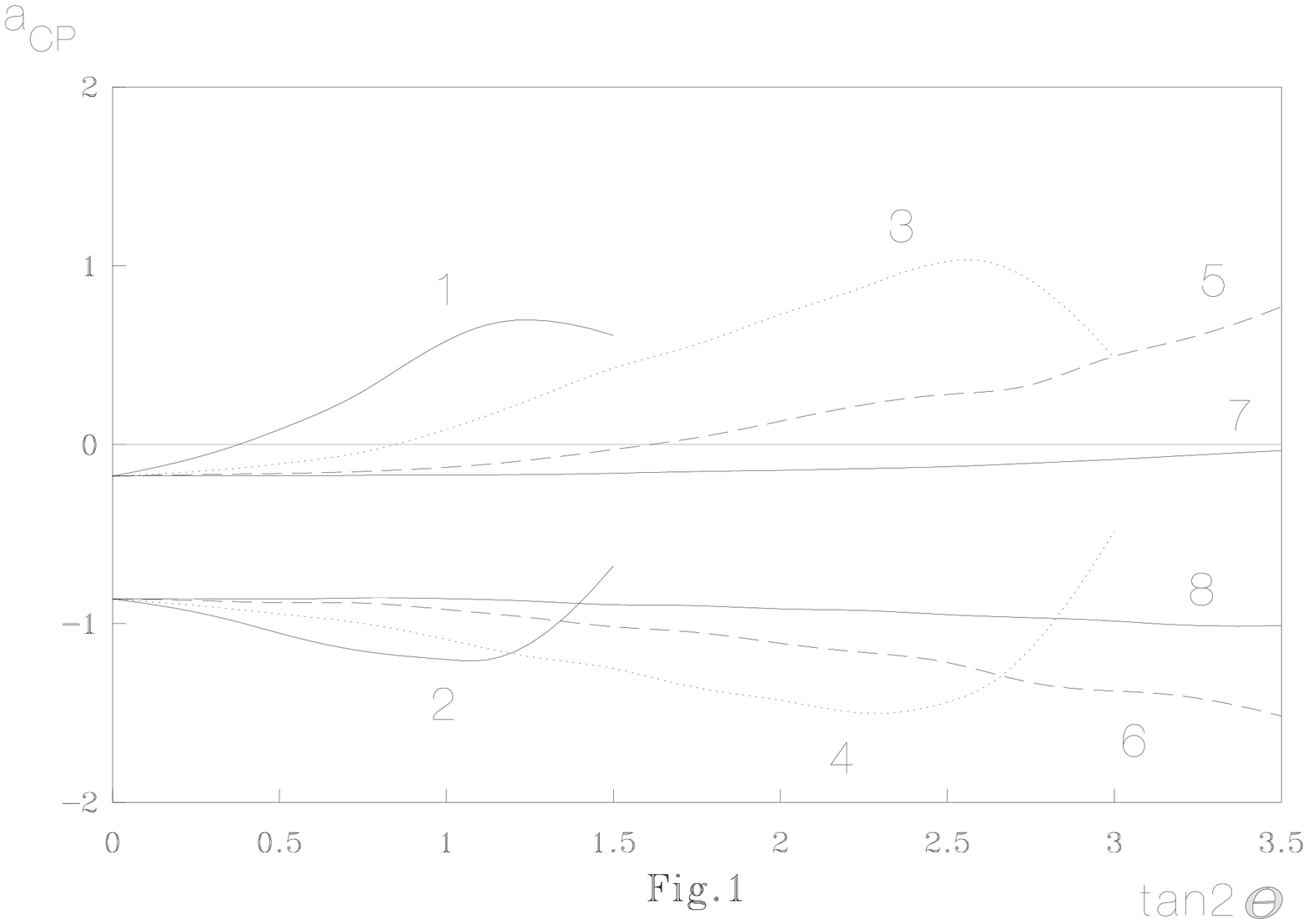}
\end{figure}
\vspace{1cm}
\begin{figure}[htb]
\epsfxsize=15cm
\epsfysize=10cm
\mbox{\hskip 0cm}\epsfbox{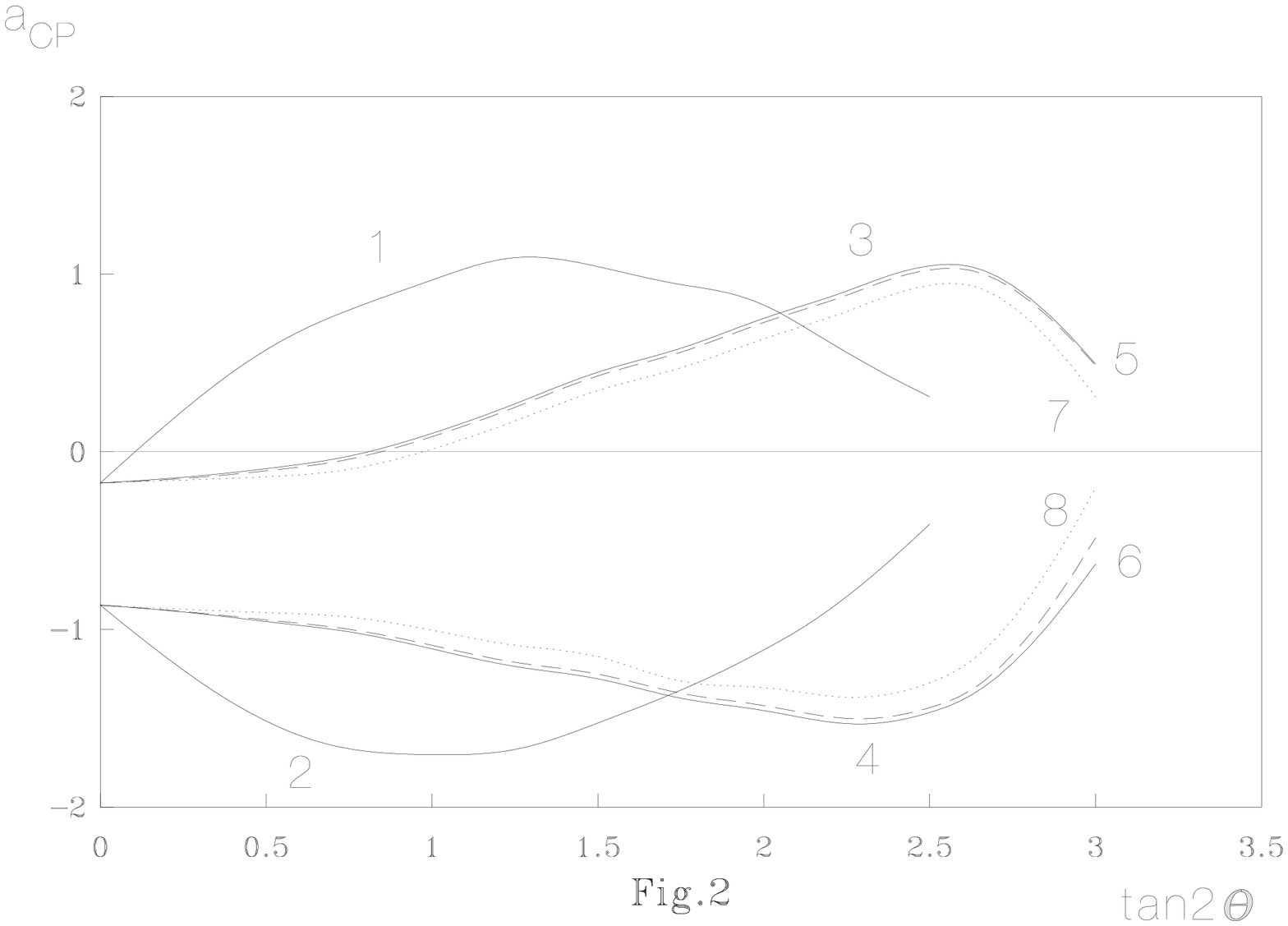}
\end{figure}
\newpage
\begin{figure}[htb]
\epsfxsize=15cm
\epsfysize=10cm
\mbox{\hskip 0cm}\epsfbox{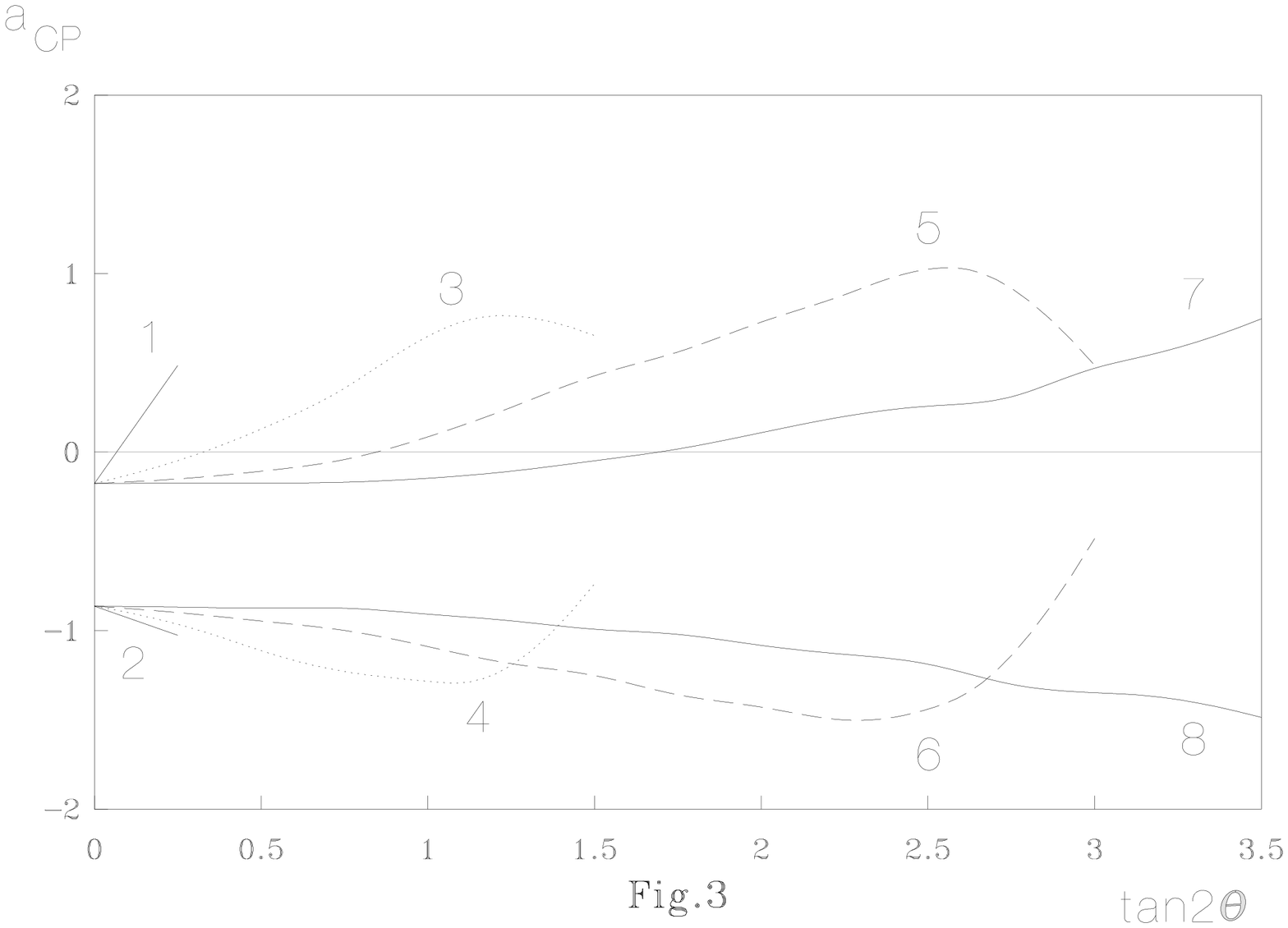}
\end{figure}
\vspace{1cm}
\begin{figure}[htb]
\epsfxsize=15cm
\epsfysize=10cm
\mbox{\hskip 0cm}\epsfbox{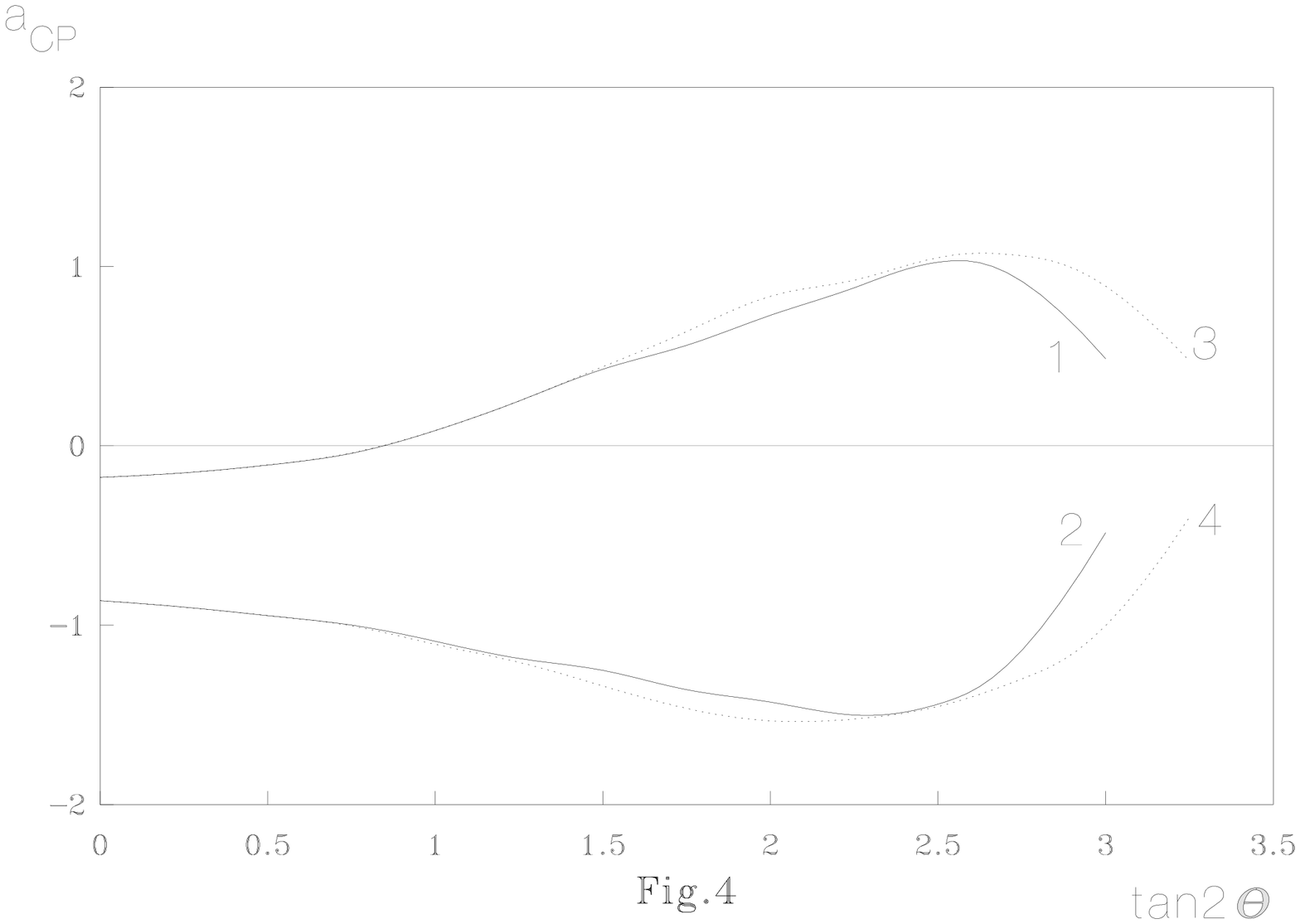}
\end{figure}
\end{document}